\documentstyle[11pt]{article}

\def \pn{\par\noindent}

\def\giorno{}

\newtheorem{theorem}{Theorem}[section]
\newtheorem{lemma}[theorem]{Lemma}
\newtheorem{corollary}[theorem]{Corollary}
\newtheorem{proposition}[theorem]{Proposition}

\begin{document}

\title{Convergence of normal form transformations: \\ The role of
symmetries}

\author{{Giampaolo Cicogna} \\
{Dipartimento di Fisica, Universit\`a di Pisa} \\
{Via Buonarroti 2, Ed.B, 56127 Pisa, Italy} \\
{cicogna@df.unipi.it} \\  \\  \\ \\
Sebastian Walcher \\
{Zentrum Mathematik},
{TU M\"unchen}\\
{80290 M\"unchen, Germany}\\
{and}\\
{Institut f\"ur Biomathematik und Biometrie, GSF-Forschungszentrum}\\
{Postfach 1129, 85758 Neuherberg, Germany}\\
{walcher@mathematik.tu-muenchen.de} }

\date\giorno

\maketitle

\begin{abstract}
We discuss the convergence problem for
coordinate transformations which take a given vector field into
Poincar\'e-Dulac normal form. We show that the presence of
linear or nonlinear Lie point symmetries can guaranteee convergence of
these normalizing transformations, in a number of scenarios.
As an application,
we consider a class of bifurcation problems.
\end{abstract}

\vfill\eject

\bigskip\medskip\pn
\parindent=20pt

\def\ds{\displaystyle}
\def\ov{\over}

\def\C{{\cal C}}
\def\G{{\cal G}}
\def \R {{\bf R}}

\def \a{\alpha}
\def \be{\beta}
\def \th{\theta}
\def \b{\beta}
\def \d{\delta}
\def \La{\Lambda}
\def \la {\lambda}
\def \l{\eta}       
\def \ep{\varepsilon}
\def \phi{\varphi}
\def \s{\sigma}
\def \om{\omega}

\def\pa{\partial}
\def\pd{\partial}
\def \xd {\dot x}
\def\grad{\nabla}
\def\Ker{{\rm Ker}}
\def\ran{{\rm Ran}}
\def\ker{{\rm Ker}}
\def\ss{\subset}
\def\sse{\subseteq}

\def \bs{\bigskip}

\def \e{{\rm e}}
\def \d{{\rm d}}
\def \ii {{\rm i}}
\def\=#1{\bar #1}
\def\~#1{\widetilde #1}
\def\.#1{\dot #1}
\def\^#1{\widehat #1}

\def\sp{\sum\nolimits'}
\def \tr {transformation}
\def \com {constant of motion}
\def \coms {constants of motion}
\def \sy {symmetry}
\def \sys {symmetries}
\def \an {analytic}
\def \co {convergen}
\def \evl {eigenvalue}
\def \bif{bifurcat}

\def \q {\quad}
\def \qq {\qquad}
\def \en {\eqno}
\def \cd {\cdot}
\def \dst {\displaystyle}


\section{Normal forms and normalizing transformations.}
Normal form theory \cite{a1,a2,b1,b2,ch,vm} (see also
\cite{CG3}, where many other references can be found) was
introduced by Poincar\'e in his thesis, as a tool to integrate
nonlinear systems. While it is now known that integration is
in general impossible, normal forms have proven to be among the most
useful tools
both in the qualitative and quantitative local analysis of
dynamical systems.

We will consider an ordinary differential equation
$$\.x\equiv{\d x\over{\d t}}=f(x)\qq\qq x=x(t)\en(1.1)$$
and the associated vector field $X_f$ (we will sometimes call
both $X_f$ and $f$ a vector field)
$$X_f\equiv \sum_{i=1}^n f_i(x){\pd\over{\pd x_i}}\qq\qq
(x\in \R^n)\en(1.2)$$
which we assume to be \an\ in a neighourhood of a stationary point $x_0$
(i.e., a point such that
$f(x_0)=0$; we can choose $x_0=0$).
The basic idea is to  introduce a  near-identity change   of
coordinates in order to eliminate nonlinear terms in the given vector
field;
certain (appropriately chosen) terms which cannot be eliminated, the
so-called ``resonant terms'',
constitute the Poincar\'e-Dulac normal form. We will elaborate details
in a moment. The coordinate \tr s are usually obtained by
means of iterative techniques: in general the normalizing \tr s are
actually formal power series \tr s, and only special conditions can
ensure their (local) \co ce and the (local) analyticity of the
normal form \cite{a1,a2,b1,b2}.

The procedure is well known; we recall some essential
points, to fix notations, and in view of the applications below (see also
\cite{tut}).

Writing the system (1.1) in the form
$$\.x=f(x)=Ax+F(x)\en(1.1')$$
we will always assume that the matrix $A=(Df)(0)$ is nonzero and
semisimple, with eigenvalues $\lambda_1,\ldots,\lambda_n$.
A normal form of $f$ will be written
$$\^f(x)=Ax+\^F(x).\en(1.3)$$
(The notation $\ \^\cd\ $ will be always reserved for vector fields in
normal form; there is no danger of confusion if $x$ will also be used to 
denote the ``new'' coordinates.)
The characterizing property of Poincar\'e-Dulac normal forms can be stated
as follows: If $e_1,\ldots,e_n$ is an eigenbasis of $A$, and
$x_1,\ldots,x_n$ denote the corresponding coordinates, then $\^F$ is a
series in those monomials
$$x_1^{m_1}\cdots x_n^{m_n}\,e_j\qq{\rm with}\qq m_1\lambda_1+\ldots
m_n\lambda_n-\lambda_j=0. \en(1.4)$$
We will say that the eigenvalues are in resonance if they satisfy a
relation as above with nonnegative integers $m_i$, $\sum m_i\geq 2$.

There is always a formal power series transformation to normal form
(but neither the normal form nor the normalizing transformation are unique
in general, see e.g. \cite{a1,bibi,b1,b2}).
A sufficient condition which ensures convergence of a normalizing
transformation is given by
the following criterion (see \cite{a1,a2} for more
details and extensions):
\begin{theorem} {\rm (Poincar\'e)}
If the eigenvalues
$\la_1,\ldots,\la_n$ of the matrix $A=(Df)(0)$  belong
to the Poincar\'e domain, i.e. if the convex hull of the points
$\la_1,\ldots,\la_n$ in the complex plane does not contain zero, then
there
is a convergent normalizing transformation (and an analytic normal form).
\end{theorem}

Building on work by Siegel and Pliss, Bruno succeeded in providing deep
insight into convergence and divergence issues.
He formulated two fundamental conditions
which ensure the convergence of a normalizing transformation (in an open
neighborhood of 0).
These are called Condition A and Condition $\om$ (see \cite{b1,b2} for
details).
In order to avoid technicalities, we state  Condition A in its
simplest (and slightly too restrictive) form:
\medskip\pn
{\bf Condition A:} A normal form $\^f$ is said to satisfy Condition
A if $\^f$ has the form $$\^f=Ax+\a(x)Ax$$
where $\a(x)$ is some scalar-valued power series
\big(with $\a(0)=0$\big).
\medskip

It follows from $\^f$ being in normal form that $X_{\^f}(\alpha)=0$; one
has also
$X_A(\alpha)=0$ (in the case of linear vector fields $f=Ax$,
we shall simply write $X_A$ instead of $X_{Ax}$).
Depending on the position of the eigenvalues
of $A$ in the complex plane, Condition A should be modified appropriately
\cite{b1,b2}. However, in all the applications
we are going to discuss,
the above formulation is sufficient. Bruno's condition improves an earlier
criterion given by Pliss \cite{pli}, which requires some formal normal
form to be linear.

The other condition is a (weak) arithmetic condition, generalizing
a criterion given earlier by Siegel, and devised to
control the appearance of small divisors.
\medskip\pn
{\bf Condition $\om$}: Let $\om_k=\min|(Q,\Lambda)-\la_j|$ for all
$j=1,\ldots,n$ and $n-$tuples
of integers $q_i\ge 0$ such that $1<\sum_{i=1}^n q_i<2^k$ and
$(Q,\Lambda)\equiv\sum_iq_i\la_i\ne\la_j$: then
$$\sum_{k=1}^\infty 2^{-k}\ln \big(\om_k^{-1}\big)<\infty$$
\medskip\pn

With these two conditions, one has  \cite{b1,b2}:

\begin{theorem} {\rm (Bruno)}
If $A=(Df)(0)$ satisfies Condition $\om$, and if $f$ can be transformed,
via a formal coordinate transformation, to some ${\^f}$ which satisfies
Condition A, then there is a convergent normalizing transformation for~$f$.
\end{theorem}

Here and in the following, ``convergent'' stands for
``convergent in some open neighbourhood of the stationary point $x_0=0$''.

Condition $\om$ is satisfied by almost all (in the Lebesgue
sense) $n-$tuples of \evl s $\la_i$. For the sake of simplicity, we
will assume that the matrix $A$ satisfies Condition $\om$,
unless we explicitly say otherwise.

Bruno also stated and proved divergence theorems which show that weaker
versions of Condition A and Condition $\omega$ are necessary to ensure
convergence in the general setting. These theorems do not, however,
address the convergence problem for a given analytic vector field,
and there are many instances of convergence where $A$
satisfies neither Condition A nor Condition $\omega$.
This is the central question to be discussed in this article, and 
symmetries
play a fundamental role here.

\section{Symmetries and invariants.}
We now start to investigate the
role of \sy\ properties of the vector field
in the convergence problem for the normalizing transformation
\cite{a2,eal,ia}
(see also \cite{CG2,CG3}). In this paper we shall consider only Lie point
\sys : given the vector field $X_f$, we will say that
the vector field
$$X_g\equiv \sum_{i=1}^n g_i(x){\pd\over{\pd x_i}} \en(2.1)$$
is an infinitesimal (Lie point) symmetry for $X_f$ if
$$ [X_f,X_g]\equiv X_fX_g-X_gX_f=0 \en(2.2)$$
or, equivalently,
$$[f,g](x)\equiv Dg(x)f(x)-Df(x)g(x) \ =\ 0, \qq {\rm all}\,\,x.
\ \en(2.2')$$
We will then also say that $X_g$ is in the centralizer of $X_f$ (or that
$g$ is in the centralizer of $f$). The  vector field $X_g$ provides the
Lie generator of a
local one-parameter group of symmetries \cite{ol,ov} of the differential
equation. We will have to deal with various classes of centralizers. By
$\C_{\rm an}(f)$ we denote the Lie algebra of local analytic centralizer
elements of $f$, while $\C_{\rm for}(f)$ is the algebra of all formal
power series vector fields commuting with $f$.

Moreover, we call a scalar-valued function $\phi$ an invariant (or a first
integral) of $f$ if $X_f(\phi)=0.$

The presence of a \sy\ can be of considerable help in the normalizing
procedure, not only in the general problem of computing normal forms
(symmetries may impose strong restrictions on the explicit expression of
normal
forms, see e.g. \cite{CG2,CG3}), but also in the study of the convergence
of a normalizing transformation. Let us recall some well known and useful
facts (see  \cite{CG2,CG3,tut,eal,ia}).

First, the ``resonant terms'', which constitute the nonlinear
part $\^F(x)$ that cannot be eliminated in the normal form
$\^f(x)=Ax+\^F(x)$,
are precisely the terms such that
$$\^F(x)\in\Ker({\rm ad}\,A)\en(2.3)$$
where ${\rm ad}\,A$ is the ``homological operator'' defined by
$${\rm ad}\,A(h)=[Ax,h]\en(2.4)$$
This characterization of the resonant terms leads to a characterization of
normal forms in terms of \sy\ properties:
\begin{proposition}
The vector field $f=Ax+F$ is in normal form if and only if $[Ax,f]=0$.
\end{proposition}
\medskip
Thus, vector fields in normal form $\^f$ always admit nontrivial commuting
vector fields $g$ (i.e. $[g,\^f]=0$, $g\notin\R f$: this follows from
the above proposition if $\^f\not= Ax$, in case $\^f=Ax$ there are
nontrivial linear commuting vector fields, e.g. $g=A^kx$, or $g=Ix$ where
$I$ is the identity matrix.)
A fundamental property of normal forms is the following \cite{w}:
\begin{proposition}
Every (infinitesimal) \sy
$$g(x)=Bx+G(x)\en(2.5)$$
of a normal form $\^f$ is also a \sy\ of the linear part $Ax$ of $\ \^f$.
Every invariant of $\ \^f$ is also an invariant of the linear part $Ax$.
\end{proposition}
\medskip

Concerning the interplay between normal forms and \sy\ properties, we have
the following first results. Given a commuting field $g(x)=Bx+G(x)$,
we will assume here, as we did for the matrix $A$,
that $B$ satisfies Condition $\om$ and that $B$ is semisimple.
\begin{proposition}
If $f$ admits a linear (infinitesimal) \sy\ $g_B(x)=Bx$, then there is a 
normal form $\^f$
which also admits this \sy . If $f$ admits a (analytic or formal)  \sy\
$g=Bx+G(x)$, then $Bx$ is a \sy\ of some normal form $\^f$, thus $\^F$
also commutes with $Bx$:
$\^F\in\Ker({\rm ad}\,A)\cap\Ker({\rm ad}\,B)$.
\end{proposition}
\noindent{\it Proof (sketch):} Transform $g$ to normal form
$\^g=Bx+\ldots.$ Then this transformation sends $f$ to some $\tilde f=
Ax+\ldots$ such that $[Bx,\~f]=0$. Now there exists a normalizing
transformation for $\tilde f$ which respects the commuting vector field
$Bx$ (see \cite{CG3,ia}).$\ \Box$

\medskip
If $g$ can be transformed to normal form by a convergent transformation
then, obviously, $f$ will be transformed to some $\tilde f$ which admits
the linear symmetry $Bx$. The next statement is more substantial.

\begin{theorem} Let the system $\.x=f(x)=Ax+F$ admit an analytic \sy\
$g(x)=Bx+G(x)$, as above. If $\Ker\,({\rm ad}\,A)\cap\Ker\,({\rm ad}\,B)$
contains only linear vector fields then there is a convergent
transformation of $f$ to normal form, and both transformed vector fields
are linear.

\end{theorem}
\smallskip
\noindent{\it Proof (sketch):} There is a formal transformation to normal
form $\^f$ that also commutes with $Bx$. The hypothesis forces the normal
form to be linear, hence $\^f=Ax$, and Condition A (even the Pliss
condition \cite{pli}) is satisfied.$\ \Box$
\medskip

\noindent {\bf Example}.
With $x\equiv(x_1,x_2,x_3)\in \R^3$, consider the following system:
$$ \begin{array}{rl} {\dot x_1} =& \ x_1 +a_1 x_1^3x_2+b_1x_1x_2^2x_3 \\
               {\dot x_2} =& -3x_2 + a_2x_1^2x_2^2+b_2x_2^3x_3 \\
               {\dot x_3} =& \ 9x_3 + a_3x_2^2x_3^2+b_3x_2^2x_3^2
\end{array} $$
where $a_i,\ b_i$ are arbitrary constants.
This system admits a linear commuting vector field
$$X_g=Bx\cdot\nabla=x_1\pa_{x_1}-2x_2\pa_{x_2}+4x_3\pa_{x_3} $$
It can be verified that all the above assumptions are satisfied,
in particular $\Ker({\rm ad}\,A)\cap\Ker({\rm ad}\,B)$ contains only linear
vector fields, although both $\Ker({\rm ad}\,A)$ and $\Ker({\rm ad}\,B)$
have infinite dimension, and neither the eigenvalues
of $A$ nor those of $B$ belong to a Poincar\'e domain. We conclude
that this vector field can be linearized by a convergent \tr .

\medskip
An especially important case occurs for $B=I$, the identity; see
\cite{BCGM}:
\begin{corollary}
A system $\.x=f=Ax+F$ can be formally linearized if and only if it
admits a formal \sy\ $g=Bx+G$ such that $B=Dg(0)=I$. If $g$ is analytic
then there is a convergent transformation to normal form.
\end{corollary}
\bigskip
As one more illustration of how normal forms may be influenced by 
symmetries,
we quote without proof a recent result \cite{gawa}:
\medskip
\begin{theorem}
Let $\cal{M}$ be the Lie algebra of a compact linear group, and suppose
that the elements of $\cal{M}$ commute with the vector field $f=Ax+F$. If
the elements of $\R\,Ax+\cal{M}$ admit no non-constant common polynomial
invariant then every normal form $\^f$ of $f$ is necessarily a polynomial.
\end{theorem}

\section{Dimension two.}
In this section we will discuss analytic two-dimensional systems
$$\.x=f(x)=Ax+F(x),$$
with $A$ having eigenvalues $\lambda_1$ and $\lambda_2$, not both zero. A
normal form of $f$ will, as usual, be denoted by $\^f$. In dimension two,
the picture is quite complete and satisfactory, and we will present the
essential ideas and sketches of proofs (see also \cite{ch}).

To start, we present an example where Condition A is violated and prove
directly (and in detail) that no convergent normalizing transformation
exists. The example is quite old, going back to Horn (1899) in a somewhat
different context, and is cited in Bruno's paper \cite{b1}. But it seems
that a complete proof of the non-existence of a convergent normalizing
transformation is not available in the literature.
\begin{proposition}
The differential system
$$ \begin{array}{rl} {\dot x_1} =& x_1^2\\
               {\dot x_2} =& x_2-x_1\end{array} \en(3.1)$$
does not admit a convergent transformation to normal form.
\end{proposition}
\smallskip{\it Proof.}
\noindent(i) An elementary computation shows that a normal form is given by
$$\^f=\left(\begin{array}{c} x_1^2\\
               x_2\end{array}\right), $$
after an intermediate step to transform $f$ to
$$\left(\begin{array}{c} x_1^2\\
               x_2-x_1^2\end{array}\right) $$
with diagonal linear part $A={\rm diag}\,(0,\,1)$.
Now assume there is a convergent normalizing transformation
$$\Phi(x)=\left(\begin{array}{c} \phi_1(x)\\
               \phi_2(x)\end{array}\right)$$
with invertible $D\Phi(0)$. This implies
$$D\Phi(x)\left(\begin{array}{c} x_1^2\\
               x_2\end{array}\right) = \left(\begin{array}{c}\phi_1^2\\

\phi_2-\phi_1\end{array}\right).$$

\noindent(ii) We claim: If $\rho\not= 0$ is a series such that $\rho(0)=0$
and $X_{\^f}(\rho)=\rho^2$ then $\rho$ is a series in $x_1$ alone. (Note 
that $\rho$ is a solution-preserving map to the one-dimensional equation 
$\dot x = x^2$.) As a
consequence, we find $x_1^2\rho^\prime(x_1)=\rho(x_1)^2$, and 
$\rho=x_1/(1+cx_1)$
follows easily. Changing the constant $c$ amounts to a coordinate 
transformation in ${\bf K}$, so we may take $c=0$ and $\rho(x)=x_1$.

To prove the claim, we show $X_A(\rho)=0$. Since $\^f$ is in normal form,
we have
$$X_{\^f}X_A(\rho)=X_AX_{\^f}(\rho)=2\rho\,X_A(\rho).$$
Now consider the Taylor expansion $\rho=\rho_r+\ldots$, with $\rho_r\not=
0$ homogeneous of degree $r$ (and $r>0$). Assume that there is a smallest
integer $q\geq 0$ such that $X_A(\rho_{r+q})\not= 0$. Then
$$2\rho\,X_A(\rho)= 2\rho_r\,X_A(\rho_{r+q})+\ldots,$$
thus the term of smallest degree in this expansion has degree $2r+q$. On
the other hand,
$$X_{\^f}X_A(\rho)= X_A^2(\rho_{r+q})+\ldots$$
forces $X_A^2(\rho_{r+q})=0$ and then $X_A(\rho_{r+q})=0$, since $A$ is
semisimple. This yields a contradiction.

\noindent(iii) Applying (ii) to the series $\Phi$, we may assume
$\phi_1(x)=x_1$.
Then $\phi_2$ must satisfy
$$x_1^2\frac{\partial{\phi_2}}{\partial{x_1}}+
x_2\frac{\partial{\phi_2}}{\partial{x_2}}=\phi_2-x_1,$$
and this forces (see \cite{b1})
$$\phi_2(x_1,0)=\sum_{k\geq 1}(k-1)!\,x_1^k.\en(3.2)$$
Therefore $\Phi$ is not convergent. $\ \Box$

\bigskip
We now present the central theorem about convergence in the
two-dimensional setting. It combines results of Markhashov \cite{Mar}, and
Bruno and Walcher \cite{BW}, and may be seen as a converse to what was 
stated
in the observation following Proposition 2.1.
\medskip

\begin{theorem}
Assume that $f$ admits a nontrivial commuting analytic vector field
$g$
. Then there is a convergent normalizing transformation for $f$.
\end{theorem}
\smallskip
\noindent{\it Sketch of proof.} Let $\lambda_1$, $\lambda_2$ be the
eigenvalues of $A$; to be specific we assume that $\lambda_1\not=0$. Let
$g=Bx+\ldots$; there is no a priori condition on $B$.

\noindent(i) If $\lambda_1$ and $\lambda_2$ are in the Poincar\'e domain
(thus, $\lambda_2/\lambda_1$ is not a negative real number) then
convergence is unproblematic by Poincar\'e's theorem.

\noindent(ii) Assume that $\lambda_2/\lambda_1<0$ and irrational. Then
only linear vector fields commute with $Ax$; in particular the normal form
is $\^f=Ax$. A formal normalizing transformation for $f$ sends $g$ to some
formal series $\tilde g= Bx+\ldots$, and $[Ax,\tilde g]=0$ forces $\tilde
g=Bx$ and $B\not=0$. Since $A$ is diagonalizable with distinct eigenvalues
and $B$ commutes with $A$, the matrices $A$ and $B$ are simultaneously
diagonalizable. If $B$ is a multiple of $A$ then $f$ is already in normal
form. Otherwise there are scalars $\sigma_1$, $\sigma_2$ such that
$\sigma_1A+\sigma_2B=I$. Now $g^*:=\sigma_1 f + \sigma_2 g$ is also a
nontrivial commuting vector field for $f$, and we can apply Corollary 2.5.

\noindent (iii) Finally, assume that $\lambda_2/\lambda_1\leq 0$ and
rational. The last argument of (ii) shows that only the case $g=\sigma
Ax+\ldots$, with a scalar $\sigma$, needs consideration. The critical point
is to show that $g^*=f+\theta g$ satisfies Condition A for some scalar
$\theta$; see \cite{BW}. So we may assume that $g=\sigma Ax+\ldots$, with
$\sigma\not= 0$ (see \cite{BW} for this), satisfies Condition A, and there
is a convergent transformation of $g$ to normal form $\^g=\sigma
Ax+\ldots$. The same transformation takes $f$ to some $\tilde f$, with
$[\^g,\tilde f]=0$, and Proposition 2.2 shows that $[\sigma Ax, \tilde
f]=0$. Hence $\tilde f$ is in normal form. $\ \Box$

\medskip

\noindent{\bf Example.} Among the equations for which the theorem is
applicable are the ``holomorphic'' 2-dimensional dynamical systems, i.e.
equations of the form
$$\.x=u(x,y) \qq\qq \.y=v(x,y) \en(3.3)$$
where, putting $z=x+iy$, the function $f(z)=u+iv$ is a holomorphic function
of the complex variable $z$. It can be verified by a simple computation
(using the Cauchy--Riemann equations) that such a system admits
the non-trivial analytic infinitesimal symmetry
$$X_g = v(x,y){\pd\over{\pd x}} - u(x,y) {\pd\over{\pd y}}\ .
\en(3.4)$$
The existence of a convergent normalizing transformation can also be seen
from consideration of the (equivalent) one-dimensional complex equation
$\dot z = f(z)=\alpha z+\ldots$
In the case ${\rm Re\ }\a=0$, these systems are of some physical
interest because they describe the ``isochronous centers'',
i.e. planar systems possessing a family of periodic orbits, of the same
period, around a stationary point.

\medskip
There is a number of detailed (and deep) results about the complex
analytic classification of germs of analytic vector fields, or rather the
associated differential forms; we mention Martinet and Ramis \cite{mara}.
Normal forms (and thus a formal classification) are but a first step
towards analytic classification. Recall the correspondence: To a vector
field $f$ one assigns the differential form $f_1\,{\rm d}x_2 - f_2\,{\rm
d}x_1$, but for any locally analytic $\sigma$ with $\sigma(0)\not=0$ the 
vector
field $\sigma \,f$ will yield an equivalent differential form, with the
same integral curves. Stated from a different perspective, the
differential equations $\dot x = f(x)$ and $ \dot x = \sigma(x)\,f(x)$
have the same local solution orbits (albeit with different
parameterizations). Thus, the convergence problem on the differential form
level amounts to convergence of a normalizing transformation for some
$\sigma \,f$. We present a result that also follows from \cite{mara}, but
we supply a different (and elementary) proof. Call a function $\rho$ an
integrating factor of $f$ if ${\rm div}\,(\rho\,f)=0$ (whence $\rho\,f$ is
locally a Hamiltonian vector field).

\begin{proposition}
Let the analytic differential equation
$\dot x = f(x)$ be given in a neighborhood of 0. Assume that
$\lambda_2/\lambda_1 = -q/p$, with positive
and relatively prime integers, or that $\lambda_1\not= 0$ and
$\lambda_2=0$.
Then there is an analytic $\sigma$, with $\sigma(0) \not= 0$,
such that $\sigma f$ admits a convergent transformation into normal form,
if and only if there is an integrating factor $\phi^{-1}$, with $\phi$
analytic in 0.
\end{proposition}
\noindent{\it Proof.}
We may assume that the formal normal form does not satisfy
Condition A. (In the case $\hat f(x)=(1+\alpha(x))Ax$ one may take 
$\phi(x)=(1+\alpha(x))x_1x_2$, so the assertion holds.)
First assume that $\lambda_2/\lambda_1 = -q/p$. There are invariant
analytic curves tangent to the eigenspaces of $A$
(see, for instance, Bibikov \cite{bibi}, Theorem 3.2, which guarantees
convergence to normal form on an invariant manifold), and we may therefore
assume that
$$f(x) =\pmatrix{x_1\beta_1(x_1,\,x_2) \cr
                                           x_2\beta_2(x_1,\,x_2) \cr},$$
with $\beta_2(0)/\beta_1(0) = -q/p$.
Let $f^{\ast} := \beta_2^{-1} f$; thus we have the orbit-equivalent
vector field
$$f^{\ast}(x)
=\pmatrix{x_1\beta_1^{\ast}(x_1,\,x_2) \cr x_2 \cr}.$$
According to \cite{wonpoi}, Theorem~2.3, since Condition A does not hold,
there is a unique integrating factor
$\bigl(x_1^{1+\ell q}x_2^{1+\ell p}\exp(\mu)\bigr)^{-1}$
(with $\ell$ determined by the
formal normal form), and \cite{wpr}, Prop. 1.1 shows that
$g:=\bigl({{x_1^{1+\ell q}x_2^{\ell p}\exp(\mu)}\atop 0}\bigr)$ satisfies
$[g,\,f^{\ast}] = \mu f^{\ast}$ for some analytic $\mu$.
On the other hand, a direct verification shows
$$[g,\,f^{\ast}]= \pmatrix{* & * \cr 0 & * \cr}\pmatrix{*\cr 0\cr}
-\pmatrix{* & * \cr 0 & 0 \cr}\pmatrix{* \cr * \cr} =
\pmatrix{* \cr 0 \cr},$$
and this forces $\mu = 0$. It now follows from Theorem 3.2 that $f^{\ast}$
admits a convergent transformation into normal form.

Next consider the case $\lambda_2 = 0$, $\lambda_1 \not= 0$. For the proof
in this case
(and any case where ${\rm div}(A)\not = 0$) one can argue like this: Given
an integrating factor, choose $g$ as in \cite{wpr}, Remark 1.2. Then
$[g,{1\over{{\rm div}(f)}} f] = 0$ can be directly verified.
For the proof of the reverse direction, see \cite{wpr}, Prop.~1.1. $\ \Box$

\section{Finite dimensional centralizers.}

Again, let $f=Ax+F$ be analytic, and $\^f=Ax+\^F$ a (formal) normal form of
$f$. In this section we will discuss the role of the centralizers ${\cal
C}_{\rm an}(f)$ (local analytic vector fields commuting with $f$) and
${\cal C}_{\rm for}(\^f)$ (formal vector fields commuting with $\^f$). In
the present section we always require that ${\cal C}_{\rm for}(\^f)$ is a
finite dimensional vector space.

Recall that $Ax\in {\cal C}_{\rm for}(\^f)$, hence the formal centralizer
of $\^f$ is not trivial. Let us first obtain a more precise description of
the linear elements in ${\cal C}_{\rm for}(\^f)$: Among the eigenvalues
$\lambda_1,\ldots,\lambda_n$ of $A$, we may assume that
$\lambda_1,\ldots,\lambda_d$ are a maximal linearly independent system
over the rational number field ${\bf Q}$. (This notation will be kept for
the remainder of the section.)

\begin{lemma}
Assume that $A$ is diagonal.

\noindent(a) There are linearly independent diagonal matrices
$A_1,\ldots,A_d$ with rational entries such that
$$A=\lambda_1A_1+\ldots+\lambda_dA_d\qq{\rm and}\qq
[A_1x,\,\^f]=\cdots=[A_dx,\,\^f]=0.$$

\noindent(b) Let the complex numbers $\sigma_1,\ldots,\sigma_d$ be
linearly independent over ${\bf Q}$. Then $Ax$ and
$(\sigma_1A_1+\ldots+\sigma_dA_d)x$ have the same nonlinear formal
centralizer elements.
\end{lemma}
\smallskip
\noindent{\it Sketch of proof.} According to the hypothesis, there are
rational numbers $\alpha_{ij}$ such that
$$\lambda_j=\sum_{i=1}^d \alpha_{ij}\lambda_i\qq(j=d+1,\ldots,n).$$
Therefore one may take
$$A_1={\rm
diag}\,(1,0,\ldots,0,\alpha_{1,d+1},\ldots,\alpha_{1,n}),\qq\ldots\qq ,$$
$$A_d={\rm diag}\,(0,\ldots,0,1,\alpha_{d,d+1},\ldots,\alpha_{d,n}).$$
Moreover, the eigenvalues $\lambda_1,\ldots,\lambda_n$ satisfy a resonance
relation
$$m_1\lambda_1+\ldots m_n\lambda_n=\lambda_j$$
if and only if the eigenvalues of $A_1,\ldots,A_d$ satisfy this relation.
$\  \Box$

\bigskip
\noindent Clearly, the formal centralizer of $\^f$ contains $\^f$ itself 
as well as
$A_1x,\ldots,A_dx$. There may be more linear centralizer elements than the
linear combinations of  these; for instance if the eigenvalues of $A$ are
all rational, and $\^f=Ax$, then $d=1$ but the space of linear centralizer
elements has dimension $\geq n$. On the other hand, if the eigenvalues of
$A$ admit resonances (thus there are normal forms different from $Ax$), and
if $\^f$ is sufficiently {\it generic} (to be precise: if sufficiently 
many of
the resonant monomials occur in $\^F$ with nonzero coefficient) then the
subspace of linear elements of ${\cal C}_{\rm for}(\^f)$ is spanned by the
$A_ix$, as can be seen from the proof of the lemma.

\bigskip
\noindent{\bf Question.} If the eigenvalues of $A$ admit resonances, is it
true for a generic $f$ that ${\cal C}_{\rm for}(\^f)$ is spanned by $\^f$ 
and
$A_1x,\ldots,A_dx$?

\bigskip
In a number of cases this is true; see \cite{Wx} for certain classes of
equations, and also the examples later in this section. No counterexample
is known, but no general proof seems to be known either.
On the other hand, it can be seen that in the Hamiltonian case, the
centralizer of the normal form is infinite dimensional. Indeed, let
$H=H_0+H_1$ be a given analytic Hamiltonian, where $H_0$ is the quadratic
part in the canonical variables, and let $\.x=J\grad_xH=f(x)=Ax+F(x)$ be 
the
associated dynamical system, with standard notations, and with
$A$ semisimple. Once in normal form, $H_0$ is a constant of motion for the
Hamiltonian $\^H$ (see \cite{a3,CG3,vm}), and therefore
it is easy to see (cf. \cite{Ci2})  that any vector field of the
form $\phi(H_0)Ax$ belongs to $\C_{\rm for}(\^f)$, where $\phi$ is any
scalar series. Actually, the
Hamiltonian structure makes the Hamiltonian case ``non-generic"; for a
review of some results
on the convergence of the normalizing transformations for this case
see e.g. \cite{CG3}.

\medskip
Now let us turn to convergence theorems involving centralizers.
The following result combines work by Markhashov \cite{Mar}, Cicogna
\cite{Ci1}, and Walcher \cite{Wx}.

\begin{theorem}
Let ${\rm dim}\,{\cal C}_{\rm for}(\^f) = k<\infty$, and assume that there
is a matrix $A^\#$, such that $A^\#x$ has the same formal centralizer
as $Ax$, and satisfies Condition $\omega$.
If ${\rm dim}\,{\cal C}_{\rm an}(f) \geq k$ then there is a convergent
transformation of $f$ to normal form.
\end{theorem}
\noindent{\it Sketch of proof.} There is a formal transformation $\Psi$ of
$f$ to normal form $\^f$; $\Psi$ sets up a correspondence
between the formal centralizers of $f$ and $\^f$. The analytic centralizer 
of
$f$ is obviously contained in the formal centralizer, and by dimension
assumptions this transformation induces a 1-1-correspondence between
${\cal C}_{\rm an}(f)$ and ${\cal C}_{\rm for}(\^f)$ (see \cite{Wx}).
In particular there
is an analytic $g=A^\#x+\ldots$ that is transformed to $A^\#x$ via $\Psi$.
Therefore $g$ satisfies the Pliss condition, and there is a convergent
transformation $\Phi$ sending $g$ to $A^\#x$. The transformation $\Phi$
sends $f$ to some analytic vector field commuting with $A^\#x$, hence
commuting with $Ax$, hence in normal form.
 $\ \Box$

\medskip
The introduction of the matrix $A^\#$ is clearly useful in the case where
$A$ does not satisfy Condition $\omega$.

\medskip
\noindent{\bf Remark.} If ${\cal C}_{\rm for}(\^f)$ is spanned by linear
vector fields and $\^f$ then the condition of the theorem is also
necessary for convergence \cite{Wx}.

\begin{corollary}
If the \evl s of $A$ are non-resonant (and pairwise different), then there
is a convergent normalizing transformation if
and only if ${\rm dim}\, \C_{\rm an}(f)=n$.
\end{corollary}
\noindent{\it Proof.} Choose $A^\#$ as a suitable diagonal matrix
satisfying Condition A. $\ \Box$

\medskip

\noindent{\bf Remark.} The results by Markhashov \cite{Mar} and Cicogna
\cite{Ci1} include more
specific hypotheses on the centralizer elements. Actually, these
conditions follow automatically from the assumptions on the
dimension  of $\C_{\rm an}(f)$ and $\C_{\rm for}(\^f)$ in Theorem 4.2.
However, the information on the
special form of the centralizer elements is sometimes valuable.

\bigskip
The reader may ask a philosophical question here: What are these results
good for? Principally, their value lies in the structural insight they
provide: Convergence and existence of symmetries are closely related.

The analytic symmetries of $f$ will generally not be accessible in an
algorithmic manner. (Recall that first order ordinary differential
equations are a big exception in that regard; see Olver \cite{ol}.) One
may find power series expansions for centralizer elements, but they pose
the same convergence problems as in the normalization of $f$.) Therefore,
one will generally have to resort to outside information when it comes to
$\C_{\rm an}(f)$. On the other hand, the formal centralizer of $\^f$ can
be explicitly determined in a number of cases, even from a finite portion
of the Taylor series; hence this part of the problem is computationally
accessible. Basically, in these cases one can give an affirmative answer
to the question we posed above.

\medskip
\noindent{\bf Example.} (See \cite{Wx} for details.) Let $A$ be diagonal
and $\lambda_1,
\ldots,\lambda_n$ be complex numbers with the following property: There are
nonnegative integers $s_1,\ldots,s_n$, not all of them zero, such that
$s_1\lambda_1+\ldots s_n\lambda_n=0$, and whenever $m_1,\ldots,m_n$ are
nonnegative integers such that $\sum m_i\lambda_i-\lambda_j=0$ for some
$j$, $1\leq j \leq n$, then $(m_1,\ldots,m_j-1,\ldots,m_n)=
k\cdot(s_1,\ldots,s_n)$ for some nonnegative integer $k$.

Then
the elements of $ {\cal C}_{\rm for}(Ax)$ are exactly the vector fields
which can be written as
$$\sum_{l\geq 0}\rho^lB_lx,\,\, {\rm with}\,\,
\rho(x):=x_1^{s_1}\cdots x_n^{s_n},\en(4.1)$$
and all $B_l$ diagonal matrices. Thus let $\^f=\sum_{l\geq 0}\rho^lC_lx$,
and suppose the genericity condition $X_{C_1}(\rho)\not= 0$ holds. The
linear vector fields in $ {\cal C}_{\rm for}(\^f)$ are then expressed by
means of diagonal matrices and admit the first integral
$\rho$. (They obviously form a vector space of dimension $n-1$.)

Now let $g=Bx+ \rho^r D_rx+\ldots$, with $D_r\not= 0$, be a nonlinear
element of  the centralizer of $\^f$ (and of $Ax$, according to Proposition
2.2). Then comparing terms of small degree shows
$$0=\left[\rho C_1x, \rho^r D_rx\right]=
\rho\cdot r\rho^{r-1}X_{C_1}(\rho)D_rx-\rho^r\cdot X_{D_r}(\rho)C_1x .
$$
Since $C_1$ and $D_r$ are diagonal, one has $X_{C_1}(\rho)=\alpha\rho$ for
some $\alpha\not= 0$ and $X_{D_r}(\rho)= \beta\rho$ for some $\beta$.
Substituting this in the above equation yields $r\alpha D_r-\beta C_1=0$,
whence $\beta
\not=0$ and $X_{D_r}=(\beta/r\alpha)\cdot X_{C_1}$. Now the equality
$X_{D_r}(\rho)=(\beta/\alpha)X_{C_1}(\rho)\not=0$ shows that $r=1$ and
$D_r=(\beta/\alpha)C_1$. This is sufficient to conclude that the formal
centralizer of $f^*$ has dimension $n$.

\medskip
More examples and a starting point for general investigations (the concept
of ``rigidity'') can be found in \cite{Wx}.

The following result, originally due to Cicogna \cite{Ci2}, may be seen as
a counterpart to Theorem 4.2.

\begin{theorem}
Assume that $ {\cal C}_{\rm for}(\^f)$ is spanned by $\^f$ and linear
vector fields. If $ {\cal C}_{\rm an}(f)$ contains a nontrivial vector
field of the form $g=\beta Ax + G$ (with some scalar $\beta$;
$g\not=\beta\,f$) then there exists a convergent normalizing
transformation for $f$.
\end{theorem}

\noindent{\it Sketch of proof:} We may assume $g= Ax+ G$ (add $f$, if
necessary). A formal normalizing transformation for $f$ takes $g$ to some
$\tilde g=Ax+\ldots$ in the formal centralizer of $\^f$. The hypothesis now
forces $\tilde g=Ax$, and we can proceed as in the proof of Theorem 4.2. $\
\Box$

\medskip\pn
{\bf Example.} Consider a 3-dimensional system
$$\.x=f(x)=Ax+F(x) \qq {\rm with}\qq A={\rm diag}(1,1,-2) \en(4.2)$$
and let $f(x)$ possess the linear
$SO_2$ \sy\ generated by $Lx$, where
$$L=\pmatrix{0 &1 & 0\cr
               -1 &0 &0 \cr
               0 &0 &0 } \en(4.3) $$
i.e. $f$ is equivariant under rotations in the plane $(x_1,x_2)$. Putting
$\rho=x_1^2+x_2^2$, this implies that $F(x)$ must be of the form
$$ F(x)=\phi_0(\rho,x_3)Ax+\phi_1(\rho,x_3)Ix+\phi_2(\rho,x_3)Lx  $$
where $I$ is the identity matrix.
If we now choose, for instance,
$$\phi_0=0,\qq \phi_1=a_1\rho x_3+a_2x_3^3,
\qq \phi_2=b\phi_1 $$
where $a_1,a_2,b$ are constants $\not=0$, then the differential equation
(4.2) also admits the non-linear \sy\
$$g(x)=\rho x_3(I+bL)x. \en(4.4)$$
The assumption $a_2\not=0$ ensures that $f$ is {\it not}
in normal form, and that the above \sy\ is not trivial.
  Now, the normal form of $f$ is of the form
$$\^f=Ax+a_1\rho x_3 (x+ b Lx)+{\rm\ higher\ order\ terms}, \en(4.5)$$
and one can show in a manner similar to the previous example that the
hypothesis of Theorem 4.4 is satisfied. We conclude that $f$ admits a
convergent normalizing transformation.

\medskip
There are several more results in this vein; we refer to
\cite{Ci1,Ci2,Wx}. To give an example (and to emphasize once again the role
of symmetries in
the convergence problem), one may assume the presence of additional
nontrivial elements in $\C_{\rm an}(f)$: this may allow, in certain 
scenarios (see \cite{Ci2}), to conclude that $\C_{\rm for}(\^f)$
is spanned by $\^f$ and linear vector fields, and then to directly apply
Theorem 4.4.

\def\s{\lambda}    

\section{An application: the ``resonant bifurcation''.}
In this section we will apply normal form
methods to show the existence of bifurcating solutions to
dynamical systems, depending on real ``control''
parameters $\l\in\R^p$. We consider a system of the form
$$\.x=f(x,\l)\equiv A(\l)x+F(x,\l) \en(5.1)$$
with $f=f(x,\l)$ analytic in a neighbourhood of $x_0=0$ and $\l_0=0$
and with $f(0,\l)=0$, and assume that, for the ``critical''
value $\l=\l_0=0$ of the parameters, the matrix
$$A_0=A(0)\en(5.2)$$
is semisimple and its eigenvalues $\la_i$ satisfy a resonance relation. We
then show the
existence -- under suitable hypotheses -- of a general class of
bifurcating solutions  in correspondence to this resonance. Details and
complete proofs can be found in \cite{Ci3}.

\begin{theorem}
Consider the equation {\rm (5.1)} and assume that for the value
$\l_0=0$ the \evl s $\s_i$ of $A_0$ are distinct, real
or  purely imaginary, and satisfy a resonance
relation. Assume also that $p=n-1$, i.e. that there are $n-1$
real parameters $\l\equiv(\l_1,\ldots,\l_{n-1})$, and finally that putting
$$a^{(i)}_k={\pd A_{ii}(\l)\over{\pd \l_k}}\Big|_{\l=0}
\qq (i=1,\ldots,n\ ;\ k=1,\ldots,n-1), $$
the $n\times n$ matrix
$$D:= \pmatrix
          { \s_1 & a^{(1)}_1 & a^{(1)}_2 & \ldots & a^{(1)}_{n-1}    \cr
            \s_2  & a^{(2)}_1 & \ldots & &     \cr
                         \ldots\cr
            \s_n & a^{(n)}_1 & \ldots & & a^{(n)}_{n-1}   } \en(5.3)$$
is not singular.
Then there is, in a neighbourhood of $x_0=0,\ \l_0=0,\ t=0$, a \bif ing
solution of the form
$$     x_i(t)=\big(\exp(\beta(\l)A_0t)\big)  x_{0i}(\l)+
              {\rm h.o.t.} \qq \ i=1,\ldots,n  \en(5.4)$$
where  $\beta(\l)$ is some function of the $\l$'s such that
$\beta(\l)\to 1$ for $\l\to 0$, and {\rm h.o.t.} stands for higher order
terms
vanishing as $\l\to 0$.
\end{theorem}
\noindent{\it Sketch of proof.} The main idea is to transform the given
system into normal form
and to enforce that the normalizing transformation is convergent, using
the convergence conditions given by Bruno \cite{b2} for normalizing
transformations on certain subsets, which extend Theorem 1.2. Once in
normal form, the equations can be easily integrated, and the solution
(5.4) is obtained coming back to the initial coordinates by means of the
inverse (convergent) transformation.
Let us illustrate the idea in the case of dimension two:
A formal normal form here will be of type
$$\^f(\eta,x)=A_0x+\alpha(\eta,x)\,A_0x+\beta(\eta,x)\,x,\en(5.5)$$
with formal series $\alpha$ and $\beta$ with zero constant term. Obviously
Condition A is satisfied if and only if $\beta(\eta,x)=0$. The decisive
point is now that the equation $\beta=0$ defines an analytic manifold (see
Bruno \cite{b2}, Theorem 2 on p.~204). The hypothesis on $D$ ensures that
this equation locally defines a function $\eta=\eta(x)$, and using this
the assertion follows. $\ \Box$

\medskip
The standard stationary \bif ion,
Hopf \bif ion, and mul\-ti\-ple periodic \bif ing solutions as well,
are particular cases of the \bif ions obtained in this way. For
instance, if $n=2$ and with imaginary \evl s,   it is easy
to see that the condition on $D$ in the theorem coincides with the
familiar ``transversality
condition'' $\d\ {\rm Re}\s(\l)/\d\l|_{\l=0}\not=0$ ensuring standard
Hopf \bif ion. (In this context one should also mention Bibikov
\cite{bibi}, \S 7, where similar arguments are used to ensure convergence
of a certain transformation to ``normal form on an invariant manifold''.)
A nontrivial example in dimension $n>2$, and
corresponding to the case of coupled oscillators with multiple frequencies,
is given by the following corollary, which immediately follows
from the theorem. For an explicit example, see \cite{Ci3}.
\begin{corollary}
With the same notations as before, let $n=4$ and
$\s_1=-\s_2=i\om_0,\ \s_3=-\s_4=mi\om_0$ (with $m=2,3,\ldots$): with
$\l\equiv(\l_1,\l_2,\l_3)\in\R^3$, let, after complexification of the
space,
$A^C(\l)$ be conjugate to the matrix $A(\l)$ such that
$A^C(0)$ is diagonal. Putting
$\displaystyle{a^{(i)}_k={\pd A^C_{ii}(\l)\over{\pd \l_k}}\Big|_{\l=0}}
\ (i=1,\ldots,4\ ;\  k=1,2,3)$, assume that
$$\det D=\det \pmatrix
          {  a^{(1)}_1  & a^{(1)}_2 &  a^{(1)}_3 & 1    \cr
             a^{(2)}_1  & \ldots  & & -1 \cr
            \ldots  & \ldots & & m \cr
             a^{(4)}_1  & \ldots &  & -m   }  \not=0  . \en(5.6)$$
Then there is a multiple-periodic bifurcating solution preserving the
frequency resonance $1:m$.
\end{corollary}

In the case of multiple \evl s of the matrix $A_0$, the
situation is a little bit more involved: for instance, the presence in this
case of first integrals of $A_0x$ of the form $\rho=x_i/x_j$, which may
enter in the expression of normal forms, prevents the
direct application of the arguments used before.

However, the presence of degenerate \evl s is typically connected to the
existence of some \sy\ property of the problem (indeed, in the absence of
symmetries, the degeneration is ``non-generic", being possibly removed by
arbitrarily small perturbations); in this situation, the arguments
used above are still applicable, with the same result \cite{Ci3}.
An example, given by  coupled
oscillators with degenerate frequencies and in the presence of a rotation
symmetry, is described in \cite{Ci3}.

\bigskip
\noindent{\bf Acknowledgement.} During the preparation of this paper the 
second named author was visiting the University of Cagliari. The 
hospitality of the Mathematics Department, and in particular of 
T.~Gramchev, is gratefully acknowledged.


\end{document}